\documentclass[aps,twocolumn,superscriptaddress,floatfix,nofootinbib]{revtex4-2}

\usepackage[utf8]{inputenc}
\usepackage{amssymb}
\usepackage[dvipsnames]{xcolor}
\usepackage{amsmath}
\usepackage{graphicx}
\usepackage{appendix}
\usepackage{algpseudocode}
\usepackage{float}
\makeatletter
\let\newfloat\newfloat@ltx
\makeatother
\usepackage{algorithm}
\usepackage{placeins}
\usepackage{bm}

\begin{document}

\global\long\def\ket#1{\left|#1\right\rangle }%

\global\long\def\bra#1{\left\langle #1\right|}%

\global\long\def\linner#1#2{\left\langle \left.#1\,\right|#2\hspace{1.2pt}\right\rangle }%

\global\long\def\rinner#1#2{\left\langle \hspace{1.2pt}#1\left|\,#2\right.\right\rangle }%


\title{Short Quantum Circuits in Reinforcement Learning Policies for the Vehicle Routing Problem}

\author{Fabio Sanches}
\thanks{These two authors contributed equally to this work}
\affiliation{QC Ware Corp., Palo Alto, CA USA}
\author{Sean Weinberg}
\thanks{These two authors contributed equally to this work}
\affiliation{QC Ware Corp., Palo Alto, CA USA}

\author{Takanori Ide}
\affiliation{
AISIN CORPORATION, Tokyo Research Center, Chiyoda-ku, Tokyo, Japan
}

\author{Kazumitsu Kamiya}
\affiliation{Aisin Technical Center of America, San Jose, CA  USA}




\date{\today}

\begin{abstract}
     Quantum computing and machine learning have potential for symbiosis. However, in addition to the hardware limitations from current devices, there are still basic issues that must be addressed before quantum circuits can usefully incorporate with current machine learning tasks. We report a new strategy for such an integration in the context of attention models used for reinforcement learning. Agents that implement attention mechanisms have successfully been applied to certain cases of combinatorial routing problems by
    first encoding nodes on a graph and then sequentially
    decoding nodes until a route is selected. 
    We demonstrate that simple quantum circuits can used in place of classical attention head layers while maintaining performance. Our method modifies the networks used in \cite{kool2018attention} by replacing key and query vectors for every node with quantum states that are entangled before being measured. The resulting hybrid classical-quantum agent is tested in the context of vehicle routing problems where its performance is competitive with the original classical approach. We regard our model as a prototype that can be scaled up and as an avenue for further study on the role of quantum computing in reinforcement learning.

\end{abstract}

\maketitle

\section{Introduction}
\label{intro}
Quantum computing technologies have improved rapidly over the past few years with a large number of groups developing commercial quantum computing hardware. These devices aim to deliver computational capacity beyond that of classical computing devices, at least for some problem classes. Google recently demonstrated the first instance of a task performed on its quantum computer \cite{2019supremacy} for which a complete classical simulation would be prohibitively costly. A similar result was recently reported in China \cite{wu2021strong}.

While such results are important milestones, there is currently no demonstration of a quantum computation providing an advantage to any problem of current commercial importance in any industry. Much progress must still be made not only on hardware development, but also on algorithms and solutions engineering.  In an attempt to make quantum algorithms more near-term, research has been dedicated to develop and analyze algorithms that would run on noisy intermediate-scale quantum (NISQ) devices which may become available over the next decade. These devices are meant to have enough quality qubits to perform computations that are beyond our ability to simulate classically, but without the ability to perform fault tolerant computation.  In addition to the development of NISQ algorithms, a lot of work has also been placed on reducing the resource requirements for existing quantum algorithms (see, for example, improvements to the amplitude estimation algorithm \cite{suzuki_amplitude, tanaka_max_likelihood,iterative_qae}). 

In this work, we focus on solving a class of vehicle routing problems, an important optimization problem with relevance across multiple industries due to its close relationship to supply chain logistics. Such problems are combinatorial in nature, but often admit good approximate solutions, at least for standard versions. However, it is often the case that in a realistic supply chain, many more considerations and constraints come into play including events that are stochastic in nature. Motivated by such additional complexity, we investigate whether one can apply techniques from reinforcement learning to solve basic versions of this problem. In particular, we closely follow the work performed in \cite{kool2018attention}. However, we replace one of the core elements of the network developed in that work---the multi-head attention mechanism---by shallow quantum circuits. We then show that such a replacement does not negatively impact the performance of the agent.

We thus demonstrate that shallow quantum circuits are sufficient to replace certain classical operations in the context of deep policy-approximating networks for reinforcement learning. From the point of view of quantum algorithms development, it showcases the promise of using quantum processors in (parts of) neural networks used in state-of-the-art machine learning techniques. Our model does not focus on proving any sort of complexity advantage for the calculation performed on the quantum computer. The goal of this work is instead to demonstrate feasibility, a necessary step on the road to achieving quantum advantage. 

There are still many challenges to be overcome before viable quantum computing solutions to machine learning or optimization problems can be developed. To obtain realize quantum speedups from quantum subroutines, it is likely fault tolerant devices will be needed. An alternative class of algorithms that attempts to overcome the fault tolerance requirement are the variational quantum algorithms, where a parameterized quantum circuit is used to evaluate a certain function, and a classical optimizer is used to update the parameters in the circuit. However, these suffer from their own issues, including the feasibility of training or picking good parameters \cite{barrenplateaus}.

The approach of this work is to replace classical computing operations by short quantum circuits, and to demonstrate the robustness of such a change in training a reinforcement learning agent to solve a version of the vehicle routing problem. The emphasis is not that these circuits have guaranteed performance advantage over other methods, but that the quantities being computed through the circuits serve as proxies for the full classical calculations, can be done with short quantum circuits, and do not hurt the performance or the ability to train the agent. This is a first step in building an increasingly quantum agent. Toward that goal, there are many other points that need to be addressed, including the ability to perform automatic differentiation. Nevertheless, to our knowledge this is the first time it has been shown that a critical portion of a well-established classical policy network (as opposed to a generic neural network) can be successfully replaced by a quantum circuit. 

\subsubsection*{Notation and Conventions}
$\mathbf{R}$ denotes the real line and $\mathbf{R}^n$ 
is the $n$-dimensional Euclidean space.
In other cases we use bold letters to 
denote one-dimensional vectors. Such vectors are
often themselves indexed
as in ``$\mathbf{h}_i$''. In this case, for each value of the index $i$, we would have a vector in $\mathbf{R}^n$ for some $n$. $i$ must not be
confused with internal index that ranges from 1 to $n$. We denote
the $L^2$ norm of $\mathbf{x} \in \mathbf{R}^n$ by $|\mathbf{x}|$.
In the context of neural networks, unless otherwise stated, feed-forward
layers as assumed to involve a linear map (with weights) as well as an 
offset (biases). Our notation and conventions for reinforcement learning
are discussed in appendix \ref{appendix:rl}.

\subsection{Vehicle Routing Problems in Industry}
\label{vrp-industry}
Optimizing supply chain efficiency has obvious importance in
a wide array industries. Many companies contend with the
problem of transporting materials between various locations
with complex constraints and costs. Vehicle routing problems,
provide mathematical formulations that capture many
of the features of these difficult logistical concerns.
Obtaining more optimal solutions to vehicle routing problems
thus translate to reduced expenses, and due to the scale
of modern supply chains, even a very small increase in 
efficiency can lead to a substantial cost reduction.

This research is, in fact, funded by Aisin Group to study
ways to make their powertrain manufacturing supply chain more efficient by leveraging quantum computing and reinforcement learning.

Aisin's powertrain systems include automatic transmissions, hybrid transmissions and manual transmissions of vehicles. These automotive components are of course composed of many parts which are assembled in Aisin factories. Trucks must regularly transport a large number of parts from many suppliers scattered across Japan to depots, warehouses, and factories.

Aisin's main goal is to minimize the delivery cost of parts from suppliers to our factories. Delivery costs are strongly tied to total truck driving distances. Therefore, minimizing the total truck driving distances is the main objective of Aisin's supply chain management. In addition, minimizing driving time naturally reduces fuel consumption and emissions from the manufacturing process.

Traditionally, an expert on the Aisin supply chain decides the truck routes relying on their intuition, situational understanding, and experience.  Relying on humans for these tasks is problematic not only because algorithms can outperform human decision making, but also because human resources are expensive. Aisin Group, for example, employs a substantial supply chain management team that must constantly monitor shipments and make difficult decisions to try to obtain an efficient supply chain. 

While Aisin group seeks to automate the solution to the supply chain problem, its large scale and combinatorial nature makes standard classical algorithms struggle to provide very good or optimal solutions to this problem.
Motivated by this difficulty, Aisin has decided to invest in exploring quantum computing solutions for such problems.

This work is a collaboration between a quantum computing software group and an a company in the automotive sector. Our findings shed light on the feasibility and current maturity of techniques in reinforcement learning and quantum computing for a important problem faced in supply chains.

\subsection{Definition of Vehicle Routing Problems}
\label{vrp-def}
A problem instance of most vehicle routing problems can be specified by a graph with $n$ nodes along with some additional ``demand information''. Typically, one of the $n$ nodes is a special node called the depot, and the remaining $n-1$ nodes are assigned positive real numbers $\{d_i\, |\, i =1, \ldots, n-1\}$ which we refer to as their initial demand. These demands must be depleted by a truck traversing nodes and returning material to the depot node. The depot node is referred do by the index $i=0$ and has no demand. The graph along with initial demands defines an instance $G = (\text{graph}, d_1, d_2, \ldots, d_{n-1})$. In the specific problem class known as the capacitated vehicle routing problem (CVRP), we additionally introduce a positive real number called the truck capacity. That capacity can be taken to be 1 without loss of generality by re-scaling all initial demands.

We always use capacity 1 by normalizing demands. In addition, in our work the agent's policy uses 2-dimensional coordinates for the nodes as inputs denoted by $\mathbf{x}_i\in \mathbf{R}^2, i=0,1,\ldots,n-1$, so we take $G$ to be defined by these coordinates, the allowed legs between coordinates, and the demand

Consider the case of the CVRP. Allowed routes always require that the ``truck'' starts and ends at the depot.  An allowed route $\xi = (\xi_1, \xi_2, \ldots, \xi_k)$ is a sequence of of nodes starting and ending at the depot for which the truck capacity is never exceeded (i.e. $\sum_{i \in \xi} d_i < 1$ ). A candidate solution is required to deliver all the demands to the depot node, which will often mean that the truck will make multiple visits to the depot. 

The vehicle routing problem that applies to our work is a slight generalization of the CVRP known as the split-delivery vehicle routing problem (SDVRP). In this case, the truck is allowed to visit nodes even when the truck's remaining capacity $c$ is less than the demand at the node $d$. The truck can pull an amount less than the demand at the node and the demand is updated accordingly. In practice the truck will always pull the largest amount that it can ($c - d$ in this case) but that is not a requirement. We do make this simplification for the algorithm considered here (following \cite{kool2018attention}).

The quality of a candidate solution can then be evaluated by a cost function $C$, which is typically chosen to be the total distance traversed or time spent by the truck. However, any real-valued function over candidate routes can in principle be used in the optimization process. For the training considered here, we stick with the Euclidean distance of the route:
\begin{equation}
    C = \sum_{j=1}^{k-1} |\mathbf{x}_{\xi_{j+1}}-  \mathbf{x}_{\xi_{j}}|.
\end{equation}

\section{Using Reinforcement Learning for VRPs}
\label{rl-vrp}
The use of Neural Networks to solve routing problems dates back to the work of Hopfield and Tank  in 1985 \cite{hopfield-tsp}, where traveling salesman problem (TSP) instances are solved. For a recent review focusing on reinforcement learning applied to combinatorial optimization problems, see \cite{rl-comb-review}. This work builds on the reinforcement learning approach for vehicle routing problems described in \cite{kool2018attention}. One nice feature of this model is the independence of the encoder on the length of the input. Such a concept is related to the Pointer Networks developed in \cite{NIPS2015_29921001}.

One previous approach to performing reinforcement learning using quantum processors is \cite{quantum-deep-q}. In that work, variational circuits are used to in the context of deep Q-learning, where the variational circuit is used to encode the action-value function. Our approach is a policy method, and it modifies the Encoder by replacing the multi-head attention mechanism by a short quantum circuit. Variational circuits have also been used to encode quantum policies in \cite{jerbi2021variational}. Our approach differs in that it does not use the variational circuits that are common in NISQ applications. Instead, we use as a starting point a policy network that has been shown to be successful to solve the vehicle routing problem classically, and replace a key component of it by a short quantum circuit. This allows us to preserve the structure of the network, and is an initial step to developing quantum algorithms for policy based methods. 

The method for combinatorial optimization for routing problems proposed in \cite{kool2018attention} is based on an attention mechanism \cite{velic}, which can be divided into two steps: in the first, the network creates an encoding of the problem instance. The result of the encoding is then used in the decoder step, which iteratively outputs a probability distribution for the next step of the route. One feature of this method is that the policy network is independent of the size of the VRP/TSP instances considered.

The encoder and decoder together then define a learned policy with parameters $\bm{\theta}$ generating a probability distribution for generating a route. The encoder first creates an embedding of the problem instance.  The decoder then uses this embedding along with an additional context vector to iteratively generate a probability distribution over the nodes. At each step, sampling from this distribution gives us the next node to be visited. This allows us to update the context vector (as well as a mask, both of which will be described in more detail below). The new context is used in the next decoding step, effectively generating a probability distribution for a route:
\begin{equation}
\label{policy-multiplicative}
    p_{\bm{\theta}} (\xi) =
        \prod_t p_{\bm{\theta}} (\xi_t \,|\, \xi_{1}, \xi_{2}, \ldots, \xi_{t-1}).
\end{equation}
In this equation,  $p_{\bm{\theta}} (\xi_t \,|\, \xi_{1}, \xi_{2}, \ldots, \xi_{t-1})$ denotes the 
probability of selecting node $\xi_t$ immediately following the prior
selection of nodes $\xi_1, \ldots \xi_{t-1}$ and $p_{\bm{\theta}} (\xi)$
is the probability of selecting the entire route $\xi$.

\subsection{Encoder Network}
\label{sec:encoder}

The idea of the encoder is to create an embedding of the problem instance.
Embedding is performed in a sequence of layers. 
First, we encode the original $n$ nodes
as vectors $\mathbf{h}_i^{(0)} \in \mathbf{R}^{d_h}$. Then, these embedded coordinates are mapped to $\mathbf{h}_i^{(1)} \in \mathbf{R}^{d_h}$ which provides a deeper embedding. This process proceeds until the final embedding is obtained.

As mentioned previously, the network is independent of the number of nodes for the instances being considered. For VRP, we start with $n-1$ factory nodes of $G$, and append to the coordinates of each factory node the initial demand associated with that node. This step is not done for the depot ($i=0$):
\begin{equation}
    \mathbf{x}_i' = [\mathbf{x}_i, d_i].
\end{equation}

The first encoding step is a learned linear projection from the instance coordinates (with dimension $d_x$) to a $d_h$-dimensional embedding space.
This mapping is performed differently for the depot ($i=0$) and the other nodes.

\begin{align}
    \mathbf{h}_i^{(0)} &= W  \mathbf{x}_i' + \mathbf{b} & i \neq 0, \\
    \mathbf{h}_0^{(0)} &= W_0 \mathbf{x}_0 + \mathbf{b}_0.
\end{align}
In these equations, $W$ is a (learned) $d_h \times d_x + 1$ matrix while
$W_0$ is a (learned) $d_h \times d_x$ matrix (the difference being due to the fact that there is no demand for the depot). $\mathbf{b}$ and $\mathbf{b}_0$ are both learned vectors in $\mathbf{R}^{d_h}$.

After this initial embedding step, the policy network consists of a sequence of multi head attention layers \cite{attention2017}. The layers also contain an additional batch normalization step \cite{ioffe15-batchnorm}, as well as skip connections \cite{skip-connection-he} and a simple fully connected feed forward layer with dropout.

\begin{align}
    \mathbf{g}_i^{(l)} &= \textrm{BN}^{l} \left (\mathbf{h}_i^{(l-1)} + 
    \textrm{MHA}_i^l(\mathbf{h}_0^{(l-1)},\ldots 
    \mathbf{h}_{n-1}^{(l-1)}) \right),
\\
     \mathbf{h}_{i}^{(l)} &= \textrm{BN}^{l} \left(
     \mathbf{g}_i^{(l)} + \textrm{FF}^l(
     \mathbf{g}_i^{(l)})
     \right).
\end{align}

In these equeations, $\textrm{BN}$ denotes batch normalization, $ \textrm{FF}$ denotes a feedforward layer, and  \textrm{MHA} is a multi-head attention layer which we explain in the next section. The layers do not share parameters with each other, and the number of attention layers (as well as the number of attention heads) are hyperparameters. 
The feed forward layer contains one hidden layer of dimension $d_{\textrm{hidden}}$, ReLU activation and dropout. 

$\textrm{MHA}$ denotes the multi-head attention mechanism. It is this portion of the policy network that we replace by a short quantum circuit, which we describe in section \ref{quantum-attention}.
Before we describe the decoder, which also uses the multi-head attention layers, we will describe the classical attention mechanism used in \cite{kool2018attention}.

\begin{figure}
\begin{centering}
\includegraphics[scale=0.6]{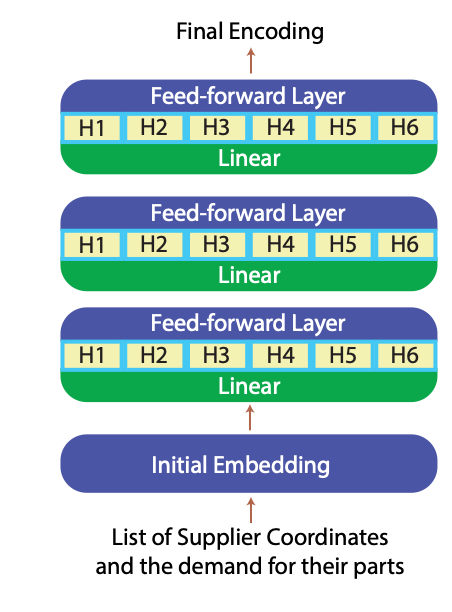}
\par\end{centering}
\caption{\label{fig:encoder-structure}Structure of the encoder in our model. This structure differs from \cite{kool2018attention} only in the heads H1, ..., H6 which are quantum rather than classical attention heads.}
\end{figure}

\subsection{Classical Multi-Head Attention Mechanism}
\label{classical-attention-section}
One simple way to understand the attention head mechanism in \cite{attention2017} is that every node is computing a compatibility with every other node. This will happen at every layer of the attention mechanism. 

To compute this compatibility, each node will have associated with it a query vector $\mathbf{q}_i$, a key vector $\mathbf{k}_i$, and a value vector $\mathbf{v}_i$. It is important that the key and query dimensions are the same $d_k = d_q$ since the compatibility will be computed as a dot product. In principle, the value vector dimension can be different, however, for our work we set $d_v=d_k$. Thus, we will use only $d_k$ to refer to the dimensionality of either of them. 

For the multi-head attention model, $M$ heads are used, each with their own linear maps determining the keys, queries, and values. For each attention head labeled by index $m\in\{1, \ldots, M\}$, the keys, queries, and values are:
\begin{align}
\label{eq:key-queries-values}
    \mathbf{q}^m_i &= Q_m \mathbf{h}_i, \\
    \mathbf{k}^m_i &= K_m \mathbf{h}_i, \\
    \mathbf{v}^m_i &= V_m \mathbf{h}_i.
\end{align}
Here, $Q_m, K_m, V_m$ are all $d_k \times d_h$ dimensional matrices. It is important to understand that in these equations, $\mathbf{h}_i$ can refer to the prior embedding at any layer, and $i$ can refer to any node. For a fixed layer and head index $m$, the same matrix $Q_m$ is applied to \emph{every node} $i$. The same is true for $K_m$ and $V_m$. All three of these matrices are learned linear maps which can vary during training. Note that we do not use biases for these maps.

The purpose of keys and queries is to compute an inter-node \emph{compatibility} $u_{ij}$. When there is an edge between nodes $i$ and $j$, we use a normalized dot product to define this:
\begin{equation}
    u^m_{ij} = \frac{\mathbf{q}^m_i \cdot \mathbf{k} ^m_j}{\sqrt{d_k}}.
\end{equation}
In the case where there is no edge on the graph between $i$ and $j$ 
(i.e. travel is not allowed between the nodes), we define
$u_{ij} = -\infty$.

The compatibility is then used to compute the attention weights using a softmax:
\begin{equation}
\label{eq:attention-weights}
    a^m_{ij} = \frac{e^{u^m_{ij}}}{\sum_{l} e^{u^m_{il}}}.
\end{equation}

The weights are finally used to compute the output of each attention head:
\begin{equation}
\label{head-output}
    \mathbf{h}_i^{\prime\,  m} = \sum_j a^m_{ij} \mathbf{v}_j^m.
\end{equation}

The last step is to then re-combine the results of each attention head. This is achieved with another set of linear maps, which maps the $d_k$ dimensional output from equation (\ref{head-output}) back into the $d_h$ dimension for the embedded nodes. Note that each head has its own linear map, and the final result is obtained by adding the mapped embeddings from each head:
\begin{equation}
\label{eq:mha}
    \textrm{MHA}_i = \sum_m A^m \mathbf{h}_i^{\prime \, m}.
\end{equation}

The result of MHA is then passed into a feed-forward layer with 1 hidden dimension and ReLU activation, followed by a 1-dimensional batch normalization with learnable parameters.

As discussed in the encoder section \ref{sec:encoder}, every MHA layer contains its own set of parameters. We have omitted the layer index in this section for readability. 

\subsection{Decoder}
\label{decoder}
For every problem instance, there is one pass through the encoder, which creates an embedding for that instance. However, the decoder is used sequentially, which each step outputting a probability distribution over the nodes of $G$, which is used to select the next location for the truck to visit. 

In the decoder, in addition to the node embeddings $\mathbf{h}_i^{(\textrm{n\_layers})}$, there is an additional context node vector. This context node plays a special role, since the attention mechanism used in the decoder only sends queries to this context node. This also means that the attention mechanism used in the decoder contains fewer computations. 

The context vector used in the VRP decoder is the following
\begin{equation}
\mathbf{c} (t) = \left [
\bar{\mathbf{h}}^{(\textrm{n\_layers})},
\mathbf{h}_{R_{t-1}}^{(\textrm{n\_layers})},
D_t
\right ].
\end{equation}
$\mathbf{c}(t)$ is therefore a vector with $2d_h + 1$ dimensions, which is updated at every timestep. 
$\bar{\mathbf{h}}^{(\textrm{n\_layers})}$ is the mean over all the embedding nodes, i.e. the output of the encoder. The second term in the concatenation is the embedding of the node selected at the previous iteration of the decoding, which is the node truck is currently at. For $t=0$, we choose $\mathbf{h}_{0}^{(\textrm{n\_layers})}$, since the truck starts at the depot. The last term is the current truck capacity, which is set to 1 initially or any time the truck visits the depot. It is updated at every step according to
\begin{equation}
    D_{t+1} = 
    \begin{cases}
        \textrm{max} \left(D_t - d_{R_t} , 0 \right)
        & R_t \neq 0
        \\
        1 & R_t = 0.
    \end{cases}
\end{equation}

In addition, the demand for the nodes are also updated at every step:
\begin{equation}
    d_{i,t+1} = 
    \begin{cases}
        \textrm{max} \left(d_{i,t} - D_t , 0 \right)
        & i = R_t
        \\
        d_{i,t} & i \neq R_t.
    \end{cases}
\end{equation}

Since the attention mechanism in the decoder only send queries from the context node, we only need one query, and keys are values are needed for every node. Again, this happens for every head separately: \begin{align}
    \mathbf{q}^m_c &= Q^{(c)}_m \mathbf{h}_i. 
\end{align}
This means that $Q^{(c)}_m$ is a $d_k \times 2d_h + 1$ dimensional matrix. The key and value computation follows equation (\ref{eq:key-queries-values}).

The compatibilities between the context and the nodes are then computed similarly, with the addition of a masking procedure. The mask $M_{i,t}$ is determined at every time step, and is the set of nodes that is note allowed to be visited. A node is included in the mask if the truck is currently located at the node (including the depot). In addition,  a node is permanently added to the mak if its demand reaches zero $d_{i,t} = 0$. And all nodes except for the depot are added to the $M_{i,t}$ at step $t$ if the truck capacity reaches 0 $D_t = 0 $, forcing the truck to return to the depot. With this in mind, the context compatibility is
\begin{equation}
    u^m_{(c)i} =
    \begin{cases}
        \frac{\mathbf{q}^m_c \cdot \mathbf{k} ^m_j}{\sqrt{d_k}} & i \notin M_{i,t}
        \\
        - \infty & i \in M_{i,t}.
    \end{cases}
\end{equation}

For every layer of the multi-head attention in the decoder the compatibilties are used to compute the outputs similar to equations (\ref{eq:attention-weights}), (\ref{head-output}),  and (\ref{eq:mha}). The decoder does not have the feed forward or batch norm layer.

The final step of the decoder involves one additional layer of the attention mechanism, with only a single attention head. This will be used to compute the probability distribution over nodes, which is in turn sampled from to determine the next node for the truck to visit. In this layer, the query and key dimensions are set to the embedding dimension ($d_k = d_h$). We use the compatibilities to compute the log probabilities in the following way:
\begin{equation}
        u^{\prime}_{(c)i} =
    \begin{cases} 10 \tanh{
    \left (
        \frac{\mathbf{q}^m_c \cdot \mathbf{k} ^m_j}{\sqrt{d_k}} \right )} & i \notin M_{i,t}
        \\
        - \infty & i \in M_{i,t}.
    \end{cases}
\end{equation}
Note that we apply the mask here as well. The factor of $10$ in front is chosen heuristically. The last step is to then compute the final probability vector:
\begin{equation}
p_i = \frac{e^{u^{\prime}_{(c)i}}}{\sum_i e^{u^{\prime}_{(c)i}}}.
\end{equation}
By sampling from this distribution, we obtain a node for the truck to designate as its next stop and append it to the route. We then proceed to update the environment, along with the quantities mentioned above.

In the next section we discuss the quantum attention head mechanism used instead of the one discussed above for the encoder. We then discuss the training and results.

\section{Quantum Attention Head Mechanism}
\label{quantum-attention}
In this section we describe the modification of the multi-head attention mechanism, replacing the compatibility calculations by expectation values of quantum states determined by a short quantum circuit.

For every key-query pair, we construct a parameterized four-qubit
quantum circuit which is measured to provide a quantum mechanical
analog of the concept of ``the compatability between the key and
query.'' The particular circuit that we describe here is a proof
of concept: it works well for the applications that we have studied
but that does not mean that it is the best choice or even the most
natural one.

To understand our circuit construction, we recall some basic aspects
of the attention mechanism of \cite{kool2018attention}. (See section \ref{rl-vrp} for more a more in-depth discussion. Here we are just recalling
the most relevant details.) In their work, given $n$ encoded nodes
in a sequence (like cities for TSP), vectors $\{k_{i}\in\mathbf{R}^{\textrm{key dimension}}\,|\,i=1,\ldots,n\}$
and $\{q_{i}\in\mathbf{R}^{\textrm{query dimension}}\,|\,i=1,\ldots,n\}$
are determined by learned parameters. The dot products $\{k_{i}\cdot q_{j}\,|\,i,j=1,\ldots,n\}$
is used as a measure of compatibility between key $i$ and query $j$.
After normalization and a softmax function, this compatibility is
used to determine a weighted message to send to the next layer.

For a quantum mechanical analog of this method, we use the following
principles:
\begin{enumerate}
\item For every encoded node $i$, rather than constructing a vector $k_{i}\in\mathbf{R}^{\textrm{key dimension}}$,
we instead construct a quantum state for two qubits $\ket{K_{i}}$.
Similarly, a quantum state for two qubits is constructed for the $i^{\textrm{th}}$
query $\ket{Q_{i}}$.
\item As with the classical mechanism, $\ket{K_{i}}$ and $\ket{Q_{i}}$
are determined by learned parameters.
\item The ``dot product'' analog should be an observable that provides
some notion of the ``similarity'' between $\ket{K_{i}}$ and $\ket{Q_{i}}$
for each pair.
\end{enumerate}

\subparagraph*{Construction of Key and Query Quantum States}

\begin{figure}
\begin{centering}
\includegraphics[scale=0.7]{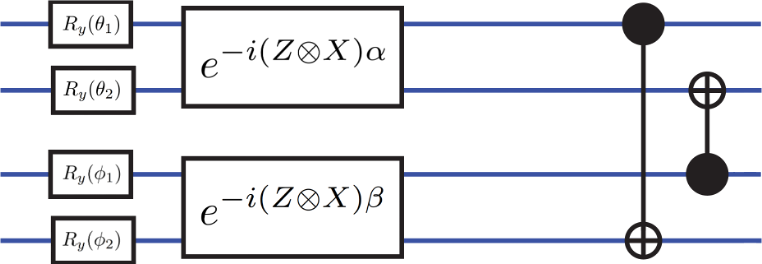}
\par\end{centering}
\caption{\label{fig:key-query-circuit}The short quantum circuit used to compute a quantum key-query compatibility.}

\end{figure}

The two-qubit quantum states $\ket{K_{i}}$ and $\ket{Q_{i}}$ are
intentionally very simple. They are both also conveniently constructed
in the same way:

\begin{align}
\ket{K_{i}} & =e^{-i\left(Z_{1}\otimes X_{2}\right)\alpha^{i}}\,R_{y1}(\theta_{1}^{i})\,R_{y2}(\theta_{2}^{i})\ket{0\,0},\\
\ket{Q_{i}} & =e^{-i\left(Z_{1}\otimes X_{2}\right)\beta^{i}}\,R_{y1}(\phi_{1}^{i})\,R_{y2}(\theta_{2}^{i})\ket{0\,0},
\end{align}
where the $\theta_{1},\theta_{2},\phi_{1},\phi_{2},\alpha,$ and $\beta$
are all $n-$dimensional parameters (one parameter for every node).
This is depicted in the left side of figure \ref{fig:key-query-circuit}
(not including the CNOT games on the right side of the figure).

After constructing a key and query quantum state for every node, the
CNOT gates shown on the right side of figure \ref{fig:key-query-circuit}
is used. This is meant to mix the key and query quantum states in
a somewhat nontrivial fashion. This simple mixing is not strictly
necessary but was chosen to test our mechanism in a context where
the key and query have entanglement structure with each other. It
is worth pointing out that this step differs from strictly mirroring
the approach taken by \cite{kool2018attention}.

The full circuit shown in figure \ref{fig:key-query-circuit} gives
rise to a key-query state $\ket{\Psi_{ij}}$ for each pair $i,j$.

\subparagraph*{Key-Query Compatibility}

To judge the ``compatibility'' between key and query quantum states,
the first obvious approach would be to consider something like the
Hilbert space inner product $\linner{K_{i}}{Q_{j}}$, but this fails
to account for the entangling CNOT gates. We could incorporate them
by computing the reduced density matrices for the key and query qubits
and using a distance metric between the density matrices. However,
these ideas do not provide any good way to be implemented on physical
quantum hardware. For that reason, we prefer to choose a definition
of compatibility that involves a physical observable. In fact, we
choose two different observables:

\begin{equation}
\label{sdots}
\mathbf{S}_{1}\cdot\mathbf{S}_{3}\textrm{ and }\mathbf{S}_{2}\cdot\mathbf{S}_{4}.
\end{equation}
These symbols require some explanation. The numbers 1, 2, 3, and 4
refer to the qubits from top to bottom in figure \ref{fig:key-query-circuit}. For $i=1,2,3, 4$, $\mathbf{S}_{i}$ denotes $\frac{1}{2}(X_i, Y_i, Z_i)$ where $X_i, Y_i,$ and $Z_i$ are the Pauli operators on qubit $i$.
The dot product notation means, e.g.
\[
\mathbf{S}_{1}\cdot\mathbf{S}_{3}=\frac{1}{4}\left(X_{1}X_{3}+Y_{1}Y_{3}+Z_{1}Z_{3}\right).
\]
This Hermitian operator, when applied to physical spin 1/2 particles,
quantifies spin-spin alignment. It is therefore not outlandish to
use it as a replacement for a classical dot product.

The actual pair of quantities that we want to pass as the output of
this entire ``layer'' of the network are the expectation values
$\left\langle \mathbf{S}_{1}\cdot\mathbf{S}_{3}\right\rangle \textrm{ and }\left\langle \mathbf{S}_{2}\cdot\mathbf{S}_{4}\right\rangle $.
From a physical hardware perspective, such expectation values can
be estimated by repeatedly sampling the quantum circuit. For a more
near-term perspective, we are interested in training to find artificial neural network (ANN) weights
and we therefore need to compute these expectation values with a quantum
simulator or with analytical formulas. The most efficient approach
for training is to first compute analytical formulas for these expectation
values in terms of the parameters $\theta_{1},\theta_{2},\phi_{1},\phi_{2},\alpha,$
and $\beta$ and to then use these formulas instead of the circuits.
This approach is excellent for such a small number of qubits, but
would fail with a larger quantum circuit. For future studies, different
circuits involving more qubits and more gates would be sensible (and
in fact, this would be the only way to discover a quantum advantage).

\section{Learning Algorithm and Methodology}
\subsection{REINFORCE Implementation}
Following \cite{kool2018attention} we implemented a variant of REINFORCE \cite{reinforce-williams}.

The REINFORCE algorithm is a policy-gradient reinforcement learning (RL) algorithm. Many
RL algorithms are based on the idea of first trying to estimate (from experience) 
the ``value'' of various actions in a given state, and then taking actions with higher
estimated value. However, policy-gradient algorithms circumvent the intermediate step
of estimating values. Instead, we assume that an agent is equipped with 
a parameterized policy $\pi$. This means that, given $\theta \in {\mathbf{R}}^D$ (where $D$ is a fixed positive integer specifying the number of parameters),
we obtain a policy $\pi(\theta)$. The policy is meant to determine the probability
that an agent takes a certain action $a$ in a state $s$. We thus write
$\pi(a\,|\,s, \theta)$ to denote the probability of taking the action $a$ in state $s$
when the policy parameters are $\theta$. The function $\pi$ is required to be differentiable (not necessarily smooth) in $\theta$.

In this context, the REINFORCE algorithm, following \cite{Sutton1998}, is given in algorithm \ref{alg:reinforce-no-baseline}.

\begin{algorithm}
\caption{REINFORCE without baseline}\label{alg:reinforce-no-baseline}
\begin{algorithmic}
\State Input: Parameterized policy $\pi$
\State Input: Initial parameter $\theta$
\Loop { (until desired performance is achieved)}:
    \State Using $\pi(\theta)$, generate episode
    \State $(s_0, a_0, r_1, \ldots, r_T) \gets$ episode
    \For {$t = 0, 1, 2, \ldots, T-1$}:
        \State $G \gets r_{t+1} + \gamma r_{t+2} + \ldots \gamma^{T-t-1} r_{T}$
        \State $\nabla J \gets \gamma^t G \, \nabla_\theta \log \left( \pi(a_t\,|\,s_t, \theta) \right)$
        \State $\theta \gets \text{Ascent}(\theta, \nabla J)$
    \EndFor
\EndLoop
\end{algorithmic}
\end{algorithm}
In this algorithm, $\gamma \in (0, 1)$ is a fixed discount factor (see appendix \ref{appendix:rl}). ``Ascent'' refers to any gradient-based ascent optimization step. This could
be gradiant ascent on $J(\theta)$ or it could be replaced by other optimization algorithms like Adam\cite{adam-diederik}. The notation we use is somewhat confusing: the usage of $\nabla J$ is just a variable name which is assigned the value on the right-hand side. Meanwhile, on the right-hand side, the symbol $\nabla_\theta$ does refer to the gradient of with respect to the parameters $\theta$ and, in particular, $\nabla J$ is a $D-$dimensional vector.  Observe also that $G$ is a constant number for the gradient computation.

Algorithm \ref{alg:reinforce-no-baseline} can learn much faster when a \emph{baseline} is
added. This essentially means that some function $b$ of states (but not of actions) is constructed with each episode and the return $G$ in the algorithm is replaced by $G - b(s)$. This algorithm still converges to the optimal policy theoretically and, with a well-chosen baseline, does so much faster. This is easiest to understand when $b(s)$ is taken to be an estimate of the return after state $s$ based on data from recent previous episodes. In this case, $G - b(s)$ being positive indicates that this episode was better than expected and thus it's sensible to increase the probability of following this sequence of actions. Meanwhile, if $G - b(s)$ is negative, the return is less than what is considered a reasonable par, and the gradient ascent would reverse and reduce the probability of taking these actions. The REINFORCE algorithm works regardless of whether or not such a baseline is used, but the learning time is dramatically reduced with a good baseline.

For our purposes we take a variant of REINFORCE adapted to our vehicle routing problem as follows. First, the
for the purposes of the algorithm we take the entire episode to be defined
by a single action. In other words, the episode is just $a_0, r_1$. The action
$a_0$ is the entire route specification $a_0 = (\xi_1, \xi_2, \ldots \xi_k)$ where each $\xi_i$ are nodes. This odd-sounding choice is sensible because with the structure of \ref{policy-multiplicative} which makes the logarithm of the policy equal to a sum over logarithms of probabilities of each action in an episode.  The reward $r_1$ is simply the negation of the route length (or time) $-L(\xi)$ which is computed by summing the appropriate distances between nodes based on a metric or on known travel times.

The second important aspect of our variant is the baseline methodology. This idea is roughly adapted directly from \cite{kool2018attention}. We maintain a ``baseline agent'' which uses the same parameterized policy but does constantly update its parameter $\theta$. Instead, the baseline agent uses an outdated parameter $\theta_{\text{BL}}$ which is occasionally updated to match the primary agent's $\theta$, but only when the agent substantially and consistently outperforms the baseline.

Our REINFORCE variant is given in algorithm  \ref{alg:reinforce-variant}. Note that this algorithm is
broken up into epochs and batches.

\begin{algorithm}
\caption{REINFORCE variant for VRP}\label{alg:reinforce-variant}
\begin{algorithmic}
\State Input: Parameterized policy $\pi$
\State Input: Integers \texttt{num\_epochs, batch\_size, batches\_per\_epoch}
\State Input: Initial parameter $\theta$
\State $\theta_{\text{BL}} \gets \theta$
\For {$e =1, \ldots,$ \texttt{num\_epochs}}:
    \For{$b =1, \ldots,$ \texttt{batches\_per\_epoch}}:
        \State $\xi \gets$  (\texttt{batch\_size} many episodes from $\pi(\theta)$)
        \State $\xi_\text{BL} \gets$  (\texttt{batch\_size} many episodes from $\pi(\theta_\text{BL})$)
        \State $\nabla J \gets \texttt{batch\_mean}\left( 
            (L(\xi) - L(\xi_\text{BL}) \nabla_\theta \log \left(\sum_{i=1}^k \pi(\xi^i, \theta) \right)
        \right)$
        \State $\theta \gets \textrm{descent}(\theta, \nabla J(\theta))$
    \EndFor
    \If {\texttt{baseline\_test()}:} \Comment{Subroutine described below}
        \State $\theta_\text{BL} \gets \theta$
    \EndIf
\EndFor
\end{algorithmic}
\end{algorithm}

One confusing part of this algorithm may be the summation $\sum_{i=1}^k \pi(\xi^i, \theta)$. To clarify, this is a sum over the probabilities computed by the encoder/decoder network at each stage of the route. $k$ refers to the number of steps in the route and the index $i$ runs over steps in the route, not over batch entries. The entire computation is performed for each batch entry and averaged over.

The \texttt{baseline\_test()} subroutine returns \texttt{true} when the policy $\pi(\theta)$ substantially outperformed the baseline policy $\pi(\theta_\text{BL})$ in recent episodes. More specifically, after each epoch we compute percentage of epochs in which the policy outperforms the baseline policy. If this percentage exceeds 50\% for 10 \emph{consecutive epochs} then we update the baseline parameters. Moreover, if the percentage exceeds 70\% for any epoch, we update the parameters. There is certainly room for experimentation with different methods here (like the one-sided T Test used in \cite{kool2018attention} but our methods were satisfactory).

\subsubsection{Training}
\label{sec:training}

For our SDVRP agent, we chose to use an embedding dimension $d_h=128$, three encoder layers with each layer having six quantum attention heads (and zero classical attention heads). The hidden feed-forward layers have dimension 128. The decoder similarly has six classical attention heads for its attention mechanisms.

To train our network, we used the Adam \cite{adam-diederik} optimizer an initial learning rate of $\eta_0 = 2^{-11}$ with exponential decay: the learning rate after $n$ episodes is given by 
\begin{equation}
    \eta_n=\alpha^{-n} \textrm{ for } n < N
\end{equation}
where $N$ is a critical number of episodes beyond which we no longer decay the learning rate. We chose $N=90$.

We trained for up to 600 epochs each of which consists of 100 batches of 128 episodes. This was performed with an NVIDIA K80 GPU. Each epoch described here took approximately 6.3 minutes to complete implying an average episode run time of  .03 seconds.
The code for this run was written in Python 3 with the PyTorch package for neural networks and CUDA integration.

Episodes were generated by taking 14 suppliers uniformly randomly distributed on a square. Their demands were randomly selected from the interval $[0, 2.5]$ for each supplier.

\section{Results}

There are three major experiments that we performed with our methodology
\begin{enumerate}
\item Training our model with synthetic data,
\item Training our model with real-world supply chain data provided by Aisin Group,
\item Implementing our attention head mechanism on Rigetti's quantum hardware.
\end{enumerate}

\onecolumngrid
\begin{center}
\begin{figure}[ht]
\includegraphics[scale=0.85]{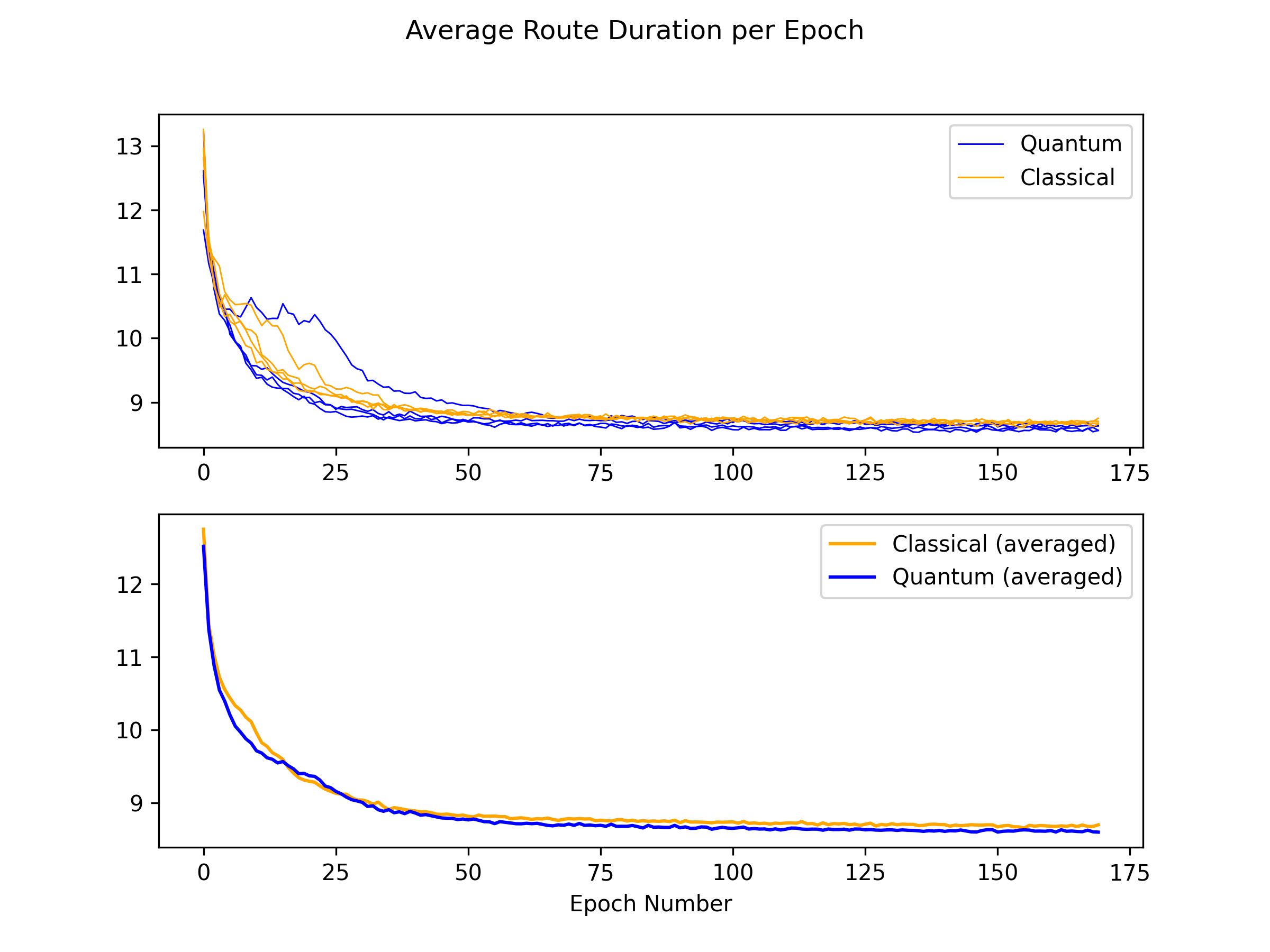}
\par
\caption{Comparison of training results for quantum and classical attention heads with synthetic data. Quantum circuits are simulated during the training process. The lower plot averages together the four quantum and four classical training runs respectively. Note that every data point in both plots is an epoch average over episodes.}
\label{fig:synthetic-training}
\end{figure}
\end{center}
\twocolumngrid

In all of these cases, we emphasize that the introduction of quantum circuits into the attention mechanism is meant as a proof-of-concept. We do not claim to have achieved a quantum advantage over classical methods. However, what we do find in all of our results is that quantum circuits \textit{can} be integrated into parameterized-policy neural networks for reinforcement learning agents \textit{while remaining competitive with conventional neural networks}. Thus, our results open the door to further investigation of hybrid quantum-classical neural networks in the reinforcement learning context.

\subsection{Synthetic Data with Simulated Quantum Circuits}
As explained in \ref{sec:training}, we constructed synthetic training data for the SDVRP by scattering 14 suppliers and one depot on the unit square. Demands were uniformly drawn integers in $\{1, 2, \ldots, 23\}$. The truck used a capacity of 1 in all cases.

The performance of our reinforcement learning agent is demonstrated in figure \ref{fig:synthetic-training}. Four runs are shown using quantum attention heads and four runs are shown using the mechanism adapted from \cite{kool2018attention}. The quantum attention heads perform as well as classical heads and in fact outperform. 

We caution that this outperformance is not a fundamentally quantum advantage. To achieve a true quantum advantage, large quantum circuits would have to be used and observables would need to be computed from them that cannot be classically simulated without a meaningful overhead. That being said, our results demonstrate that our methodology is robust and competitive: adding quantum circuits into the attention head of an encoder does not reduce performance in any obvious way and may offer a benefit. Our data suggests the need for investigation on the question of whether or not nontrivial quantum circuits in policy-approximating neural networks can provide a performance advantage.


\subsection{Performance with Aisin Supply Chain Data}

To test the resilience of our model against highly idiosyncratic real-world data, we collaborated with Aisin, a Japanese corporation that manufactures automotive components.

As discussed in section \ref{vrp-industry}, Aisin constructs powertrains using parts from a large number of suppliers scattered across Japan which must be brought to factories located in Anjo city in Aichi prefecture. These parts are carried by a fleet of trucks which represent a substantial financial cost. Improved efficiency in the logistics of part delivery constitutes a significant cost-reduction to this company.

\begin{figure}[ht]
\begin{centering}
\includegraphics[scale=0.61]{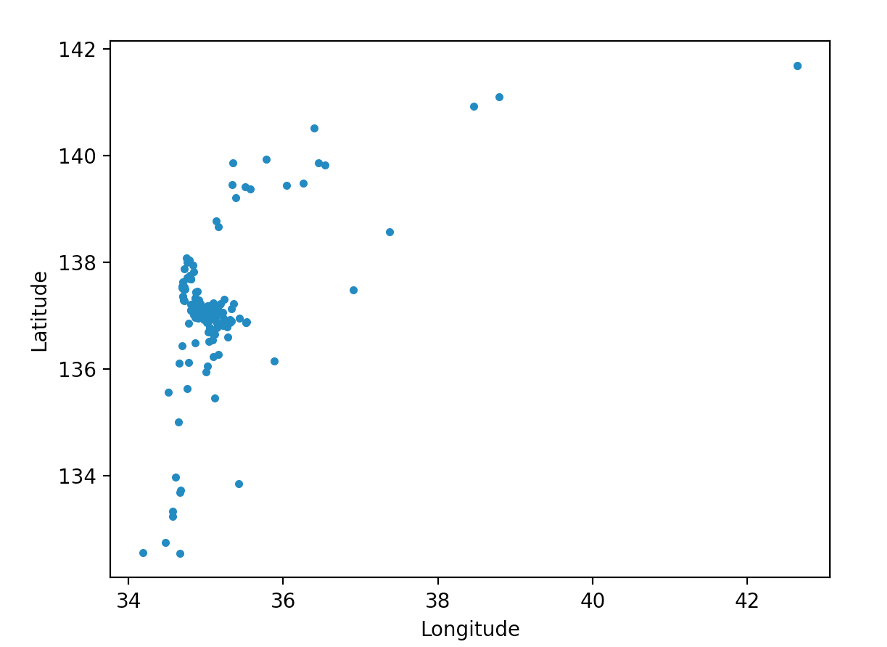}
\par\end{centering}
\caption{Locations of the 288 suppliers in the Aisin supply chain data. These suppliers are scattered across Japan. Note that the largest cluster is in the vicinity of Anjō where the depot is located.}
\label{fig:aisin-suppliers}
\end{figure}

The data set we used was collected from one day of current truck routing used by Aisin for the manufacturing of powertrains. The data set consists of the location of one Aisin factory in Anjo along with the locations of 288 suppliers across Honshu. The data set also contains information about which parts must be delivered from which suppliers and in what quantities.
The supplier distribution is shown in figure \ref{fig:aisin-suppliers}.

Because our methodology only applies to one truck, it was necessary to rescale the problem size. We randomly selected only 12 suppliers for each episode. This can be thought of as a component of a multi-truck algorithm where delivery tasks are first decomposed and split between trucks and then every truck carries out episodes in their jurisdiction based on our reinforcement learning agent.

The process of selecting 12 suppliers at a time from the data set can be thought of as sampling initial states from a probability distribution. Technically, there are only finitely many initial states which runs the risk of over-fitting. This concern is reduced by the enormous number of ways of selecting 12 suppliers from 288 choices, ensuring with near certainty that every episode involves a unique collection of nodes.

In addition, we point out that we are reporting only training progress, and do not test trained models on instances sampled from a different probability distribution. It would be interesting to understand the the ability of the model constructed here to generalize. However, for the purposes of developing a model that is useful in a particular supply chain, it suffices that it performs well for instances generated in that particular supply-chain. Therefore, we do not investigate generalizability of our model in this work. 

\begin{center}
\begin{figure}[ht]
\includegraphics[scale=0.35]{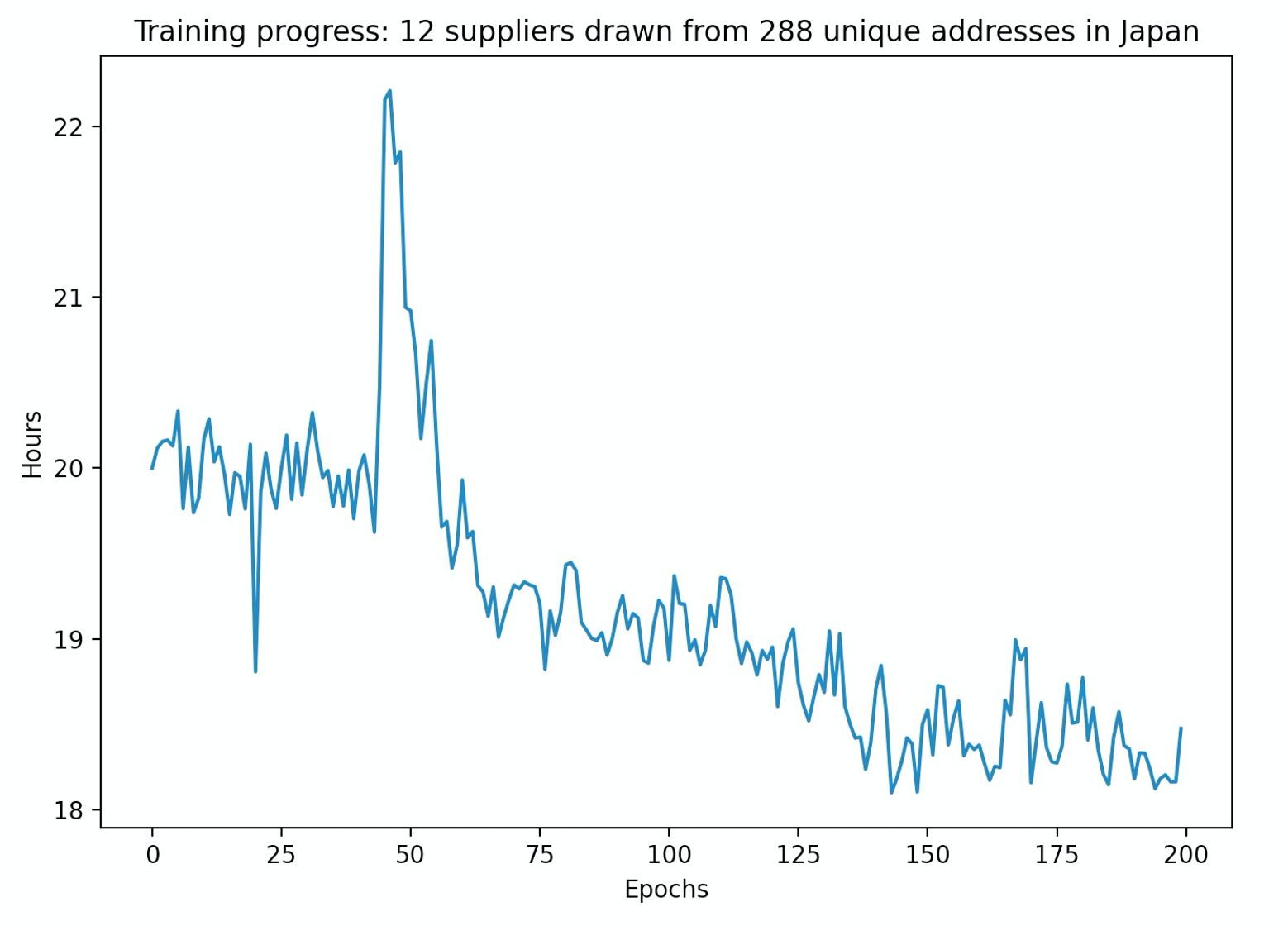}
\par
\caption{Performance of our quantum attention head model on Aisin Group powertrain supply chain data. The performance is less consistent than that of synthetic data due to the awkward distribution of suppliers.}
\label{fig:aisin-training}
\end{figure}
\end{center}

The performance of our quantum attention head model when trained on this data set is shown in
figure \ref{fig:aisin-training}. The performance improves during training as desired although the progress is somewhat unstable. These results are similar in the classical attention head case. The data used here is not particularly well-suited for our model (or that of \cite{kool2018attention} for a number of reasons. Most importantly, many legs are located over extremely long distances (Hiroshima to Anjo for instance) which result in difficulty with training. As a result, it would be more promising to assign these long haul trips by hand to certain trucks and then to use our algorithm to optimize clusters with shorter travel times. 

\subsubsection{Implementation on Rigetti}
\label{sec:rigetti}
The experiments described above were all performed by
simulating quantum circuits. We used analytical formulas for the observable
expectation values that arise with the quantum attention head mechanism,
and we used those formulas to construct functions that can be implemented
as layers of a network. This approach allowed us to test the quantum
attention head mechanism with a standard machine learning package
(PyTorch) and train and run on GPU hardware. However, there is no
point in developing a neural network with quantum circuits without
considering how an actual hardware run can play out.

\begin{figure}
\begin{centering}
\includegraphics[scale=0.3]{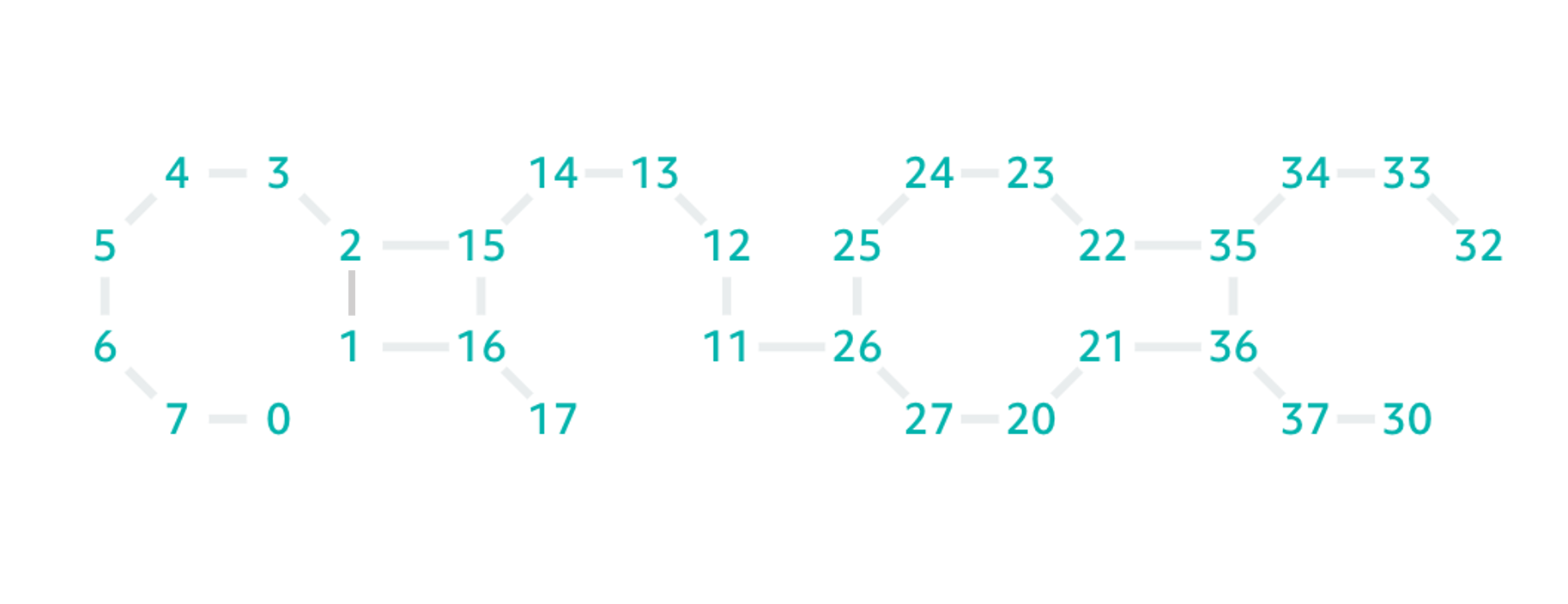}
\par\end{centering}
\caption{\label{fig:rigetti-connectivity}The connectivity of the Rigetti Aspen-8 hardware at the time of our experimentation.}

\end{figure}

For these purposes, we designed and performed an experiment on a quantum
computer developed by Rigetti Computing. Their hardware, the Aspen-8
chip had 30 qubits at the time of our experiment. Like all current
quantum hardware, that built by Rigetti suffers from gate implementation
error, readout error, and decoherence. The chip had a reported $\text{C}Z$-fidelity
of $95.55\%\pm0.15\%$. Additional summary parameters for Rigetti's Aspen-8 chip can be found on Rigetti's website \cite{rigetti_web}. Another important issue is that the connectivity
graph, shown in figure \ref{fig:rigetti-connectivity}, is not all-to-all
connected. This means that SWAP gates are necessary to perform certain
operations between distant qubits.

Given access to fault-tolerant and high clock-speed hardware, we would
perform an experiment involving training the network with the REINFORCE
algorithm. However, this would require an currently infeasible number
of calls to the hardware. Thus, we instead focus on performing a single
episode. We used parameters learned from training on a simulator and
we use those parameters to run a single small SDVRP instance. The
instance we chose consists of only 5 nodes including the depot, and
the demands are such that all nodes can be satisfied with only two
stops at the depot.

Even for a single episode, a very large number of calls must be made
to quantum hardware (discussed in further detail below). For this
reason, even with a simple instance, it was financially infeasible
to perform more than one episode for our experiment. Because of this
limitation, the actual routing result should not be taken as statistically
meaningful: it is only a single sample. Nonetheless, our implementation
is a proof-of-concept for a quantum hardware run of an episode performed
by a reinforcement learning agent with quantum circuits built into
its network on a physical quantum computer. This is a stepping-stone
toward future quantum implementations of reinforcement learning methods,
so we will focus on technical lessons learned from the experiment.

Consider the calls that must be made to hardware. For every attention
head in each encoder layer, we must construct a four-qubit circuit
for every key-query pair. Since the instance we used has five nodes,
there are $5^{2}=25$ key-query pairs (self-interactions are included).
With six attention heads per layer and three encoder layers, there
are a total of
\[
\underset{\text{node pairs}}{25}\times\underset{\text{heads per layer}}{6}\times\underset{\text{encoder layers}}{3}=450\text{ four-qubit circuits}
\]

For every circuit, it is necessary to measure six observables: 
\begin{align*}
X_{1}X_{3},\,Y_{1}Y_{3},\,Z_{1}Z_{3}\\
X_{2}X_{4},\,Y_{2}Y_{4},\,Z_{2}Z_{4}
\end{align*}
There is no obstruction to measuring, e.g., $X_{1}X_{3}$ and $X_{2}X_{4}$
in the same shot since they involve disjoint qubits. It is necessary,
however, to make three separate runs for the $X,Y,$ and $Z$ observables.
This suggests a total of 1350 calls to the Rigetti hardware (each
of which with a sufficient number of shots to collect data on expectation
values). Given that each call can take $\sim15$ sec, this is a multi-hour
run on a device with high cost and stringent reservation limitations.
To mitigate this issue, we took advantage of the fact that the chip
has 30 qubits but we only require 4 qubit circuits. This allows for
fitting six or seven circuits on the chip in parallel.

\subparagraph*{Elimination of SWAP Gates}

The connectivity graph shown in figure\ref{fig:rigetti-connectivity}
indicates that it is not possible to fill the chip with four qubit
circuits without introducing SWAP gates. This can be seen by considering
that the circuit of figure \ref{fig:key-query-circuit} involves two-qubit
gates acting on qubits 1-2, 3-4, 1-4, and 2-3. This is a substantial
concern: every additional gate results in greater error. However,
there is an approach to eliminating SWAP gates at the cost of adding
extra qubits to the circuit.

\onecolumngrid
\begin{center}
\begin{figure}
\includegraphics[scale=0.50]{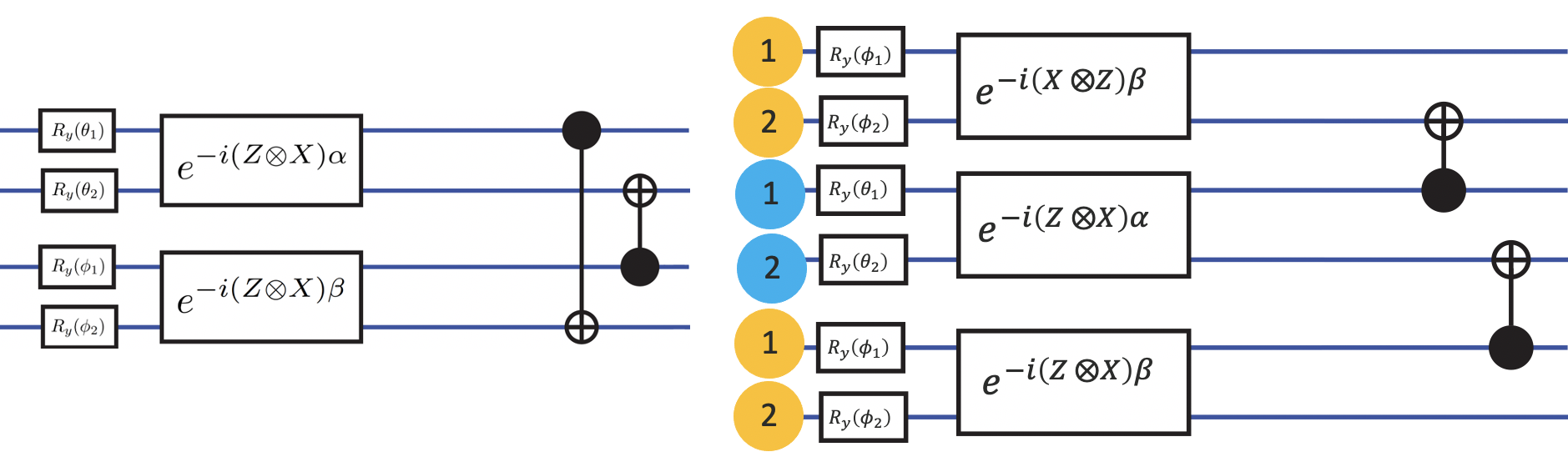}

\caption{\label{fig:embed} Our qubit mapping methodology for implementing the quantum attention head short circuit (left) on physical qubits on a line. By taking advantage of the larger number of qubits, we are able to remove error associated with SWAP gates by double the number of query qubits. In this figure, blue qubits are keys and yellow qubits are (redundantly encoded) queries.}

\end{figure}
\end{center}

\begin{figure}[ht]
\begin{centering}
\includegraphics[scale=0.39]{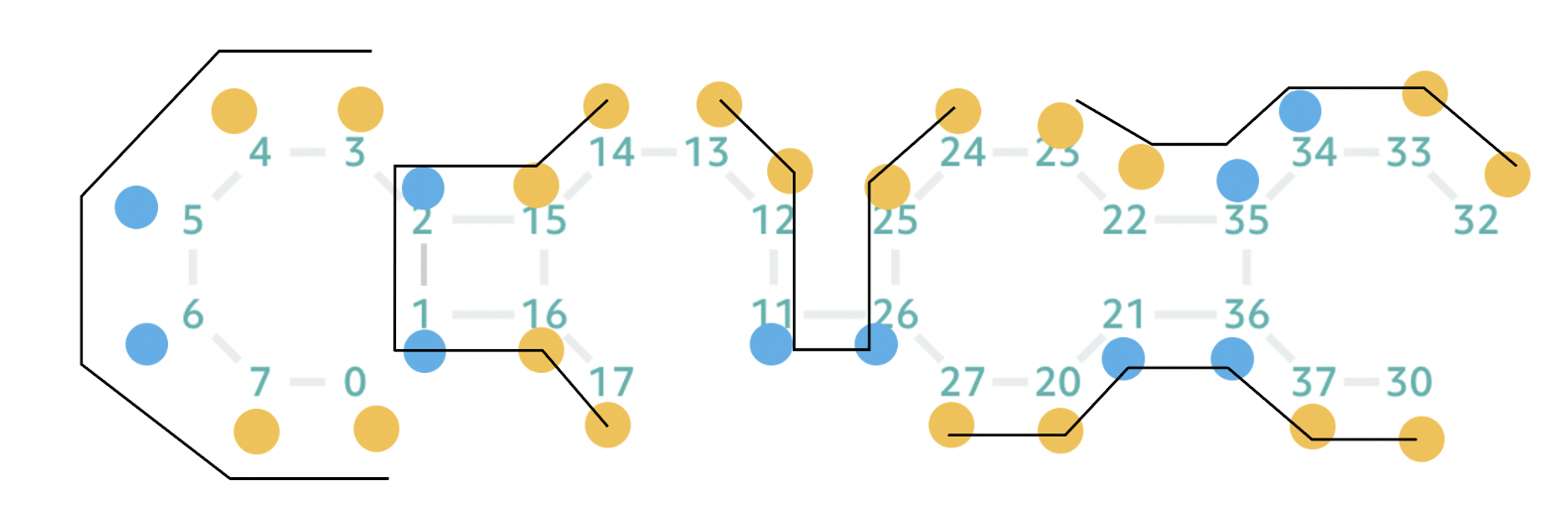}
\par\end{centering}
\caption{\label{fig:rigetti-chosen-qubits} Selection of qubits on the Rigetti Aspen-8 chip for five attention heads in parallel. The  groups of qubits that are indicated with black curves correspond directly to the construction shown in \ref{fig:embed}}
\end{figure}

\twocolumngrid
Consider the two diagrams in figure \ref{fig:embed}. The diagram
on the left is the standard quantum attention circuit and the diagram
on the right is an augmented circuit that we actually used in our
experiment instead of the one on the left. To understand this figure,
let's label the qubits in the left diagram, from top to bottom, 
\[
q_{1}^{\text{key}},q_{2}^{\text{key}}q_{1}^{\text{query}},q_{2}^{\text{query}}
\]
 and label the qubits in the right diagram, from top to bottom, 
\[
\tilde{u}_{1}^{\text{query}},u_{2}^{\text{query}},u_{1}^{\text{key}},u_{2}^{\text{key}},u_{1}^{\text{query}},\tilde{u}_{2}^{\text{query}}.
\]
This labeling will guarantee expectation values of observables on
the original qubits $q_{1}^{\text{key}},q_{2}^{\text{key}}q_{1}^{\text{query}},q_{2}^{\text{query}}$
will be the same as the corresponding measurements with the identification
\begin{align*}
q_{1}^{\text{key}} & \to u_{1}^{\text{key}}\\
q_{2}^{\text{key}} & \to u_{2}^{\text{key}}\\
q_{1}^{\text{query}} & \to u_{1}^{\text{query}}\\
q_{2}^{\text{query}} & \to u_{2}^{\text{query}}.
\end{align*}
Note that for this embedding, the $\tilde{u}$ qubits are auxillary.

Figure \ref{fig:embed} showns that there will be no need for any
SWAP gates if the 6-qubit effective circuit is used on a graph with
connectivity structure of the form
\[
\tilde{u}_{1}^{\text{query}}\leftrightarrow u_{2}^{\text{query}}\leftrightarrow u_{1}^{\text{key}}\leftrightarrow u_{2}^{\text{key}}\leftrightarrow u_{1}^{\text{query}}\leftrightarrow\tilde{u}_{2}^{\text{query}}
\]
and in fact, this line-like structure is easily found throughout the
graph in figure \ref{fig:rigetti-connectivity}. The particular embedding
that we chose to use is shown in figure \ref{fig:rigetti-chosen-qubits}.
Notice that we are able to fit five simultaneous circuits which substantially
decreases the run time.

\subparagraph*{Episode Run}

\begin{figure}[ht]
\centering{}\includegraphics[scale=0.45]{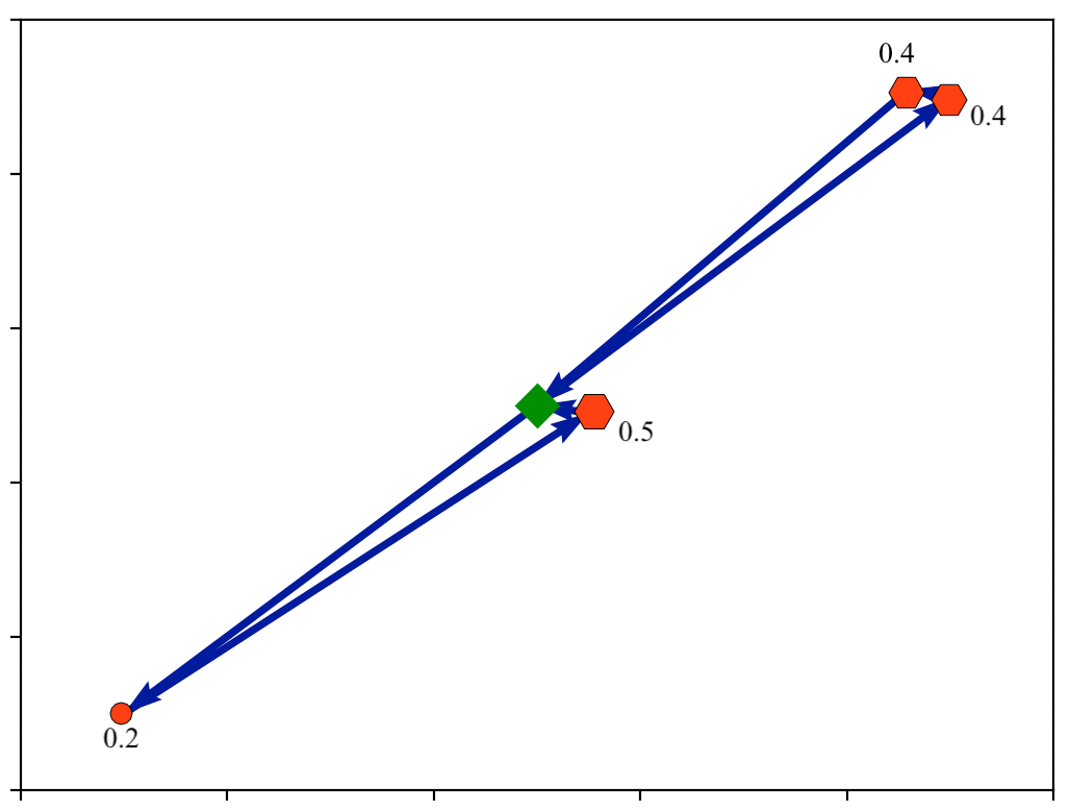}\caption{\label{fig:episode} Single VRP episode performed on Rigetti Aspen-8. The green diamond is the depot and the numbers near nodes are demands. The truck has capacity 1. We emphasize that this single episode is only meant as a proof-of concept for the feasiblity of running our model on hardware. The fact that the agent selected the optimal route is not statistically meaningful since cost constraints limited our ability to repeat the experiment.}
\end{figure}

As discussed above, running our algorithm on the Aspen-8 chip is only
barely feasible. The single episode that we succeeded in running is
shown in figure \ref{fig:episode}. The green diamond is the central
depot and the other nodes are suppliers with demands shown. Truck
capacity is 1. We emphasize that the fact that this particular run
happens to be optimal should not be taken as a statistically meaningful
statement: this could have easily occurred by random chance. The important
result here is not the routing itself but rather the fact that an
episode of the VRP was successfully performed by a ``quantum RL agent''
on hardware.

As explained above, there are 450 four-qubit circuits that must be
constructed for encoding. Since there are three observables that must
be separately observed, this would correspond to 1350 calls to the
chip (each one with 500 shots to gather statistics). However, with
the parallel approach in figure \ref{fig:rigetti-chosen-qubits}, the
actual number of call was only 270. This entire process took approximately
1.5 hours to complete. The majority of this time seems to be the delay
associated with sending different API calls to the Aspen-8 chip with
different circuit parameters. This time delay is very large compared
with the time between shots for a single API call, so there is reason
to be optimistic that such a prohibitively long time delay for circuits
in iterative algorithms with changing parameters will not be such
a difficulty in the near-term.

\section{Conclusion}
The potential for a symbiosis between quantum computing and machine
learning is a tantalizing but still undeveloped idea. There are numerous
quantum algorithms for machine learning tasks and there are also approaches
to using machine learning to improve quantum algorithms. Our focus has been
on something a bit different: implementing quantum circuits as a replacement
for a structure in a classical neural network that plays a roll in the parameterized
policy of a reinforcement learning agent.

Our findings, as summarized in figure \ref{fig:synthetic-training}, demonstrate
that our model is robust despite the nontraditional usage of quantum circuits.
The circuits do not harm the final performance of the trained agent and
the trained model with quantum attention heads actually outperformed our
implementation of the model developed in \cite{kool2018attention}.

Our model is a proof-of-concept. It does not and is not meant to represent a true quantum advantage
over classical methods. The circuits we used are too small to do so and, in particular,
the parameterized quantum circuits can be efficiently classically simulated.
Nonetheless, the observables used are nonlinear functions that would not
traditionally be considered in machine learning models. The fact that these
nonlinear functions are not harmful to the policy performance is intrinsically scientifically
interesting and suggest new creative layers that could be used in machine learning models.

In addition, we implemented the technique on real quantum computing hardware, for a toy model as described in section \ref{sec:rigetti}. Even with the  short quantum circuits with 4 qubits, demonstrating the forward pass of the policy on Rigetti's Aspen-8 chip required a redundant embedding to avoid SWAPs needed from the limited qubit connectivity. In addition, the large number of $\mathbf{S} \cdot \mathbf{S}$ (from  \ref{sdots}), required parallelizing the quantum attention circuits across the chip. Both of these techniques could be useful across different applications, especially on NISQ devices.

The major question that our research now asks is: what would happen if a reinforcement learning agent
were equipped with a policy that involved deep quantum circuits with
many qubits and nontrivial entangling structure? This cannot be probed with current NISQ
hardware, but can perhaps be investigated with high-performance quantum circuit simulators.

\bibliographystyle{unsrt}
\bibliography{bibliography}

\begin{appendix}
 \section{Review of Reinforcement Learning}
 \label{appendix:rl}
 
Reinforcement learning (RL) concerns a class of machine learning algorithms
for training agents in Markov decision processes to perform well.
A Markov decision process (MDP) describes a stochastic environment
in which an agent lives. The agent is in a particular state, and the
agent can perform actions in its state. The actions result in a change
in the environment state and a reward from the environment. Suppose
that the agent begins in the state $s_{0}$ and performs the action
$a_{0}$. The environment ``receives'' the action from the agent
and ``returns'' a reward $r_{1}\in\mathbf{R}$ and a new state $s_{1}$
(as far as the agent is concerned, the reward and new state were processed
by a black box). Continuing leads to a trajectory which means
a sequence 
\[
s_{0},\enskip a_{0},\enskip r_{1},\enskip s_{1},\enskip a_{1},\enskip r_{2},\enskip s_{2},\ldots
\]

in more detail, an MDP consists of the following: 
\begin{enumerate}
\item A state space: A finite set $\Sigma$ called the state space. Elements
of $\Sigma$ are called states. (The idea is that an element
of $\Sigma$ is a state that the agent can be in. Note that
elements of $\Sigma$ may or may not completely describe the environment.) 
\item Available actions for every state: For every state $s\in\Sigma$,
we have a finite set $B_{s}$ of objects called available actions.
($B_{s}$ is the collection of actions that the agent can perform
in the state $s$.) 
\item A reward set: a finite set $K$ of real numbers. (These are the possible
rewards--it's helpful to introduce $K$ because it's useful and generally
harmless to assume that there are only finitely many possible rewards.
In many MDPs, there are only two possible rewards--a ``good'' and
a ``bad''.) 
\item A fixed stochastic dynamical evolution law: given $s,s'\in\Sigma$,
$a\in B_{s}$ and $r\in K$, we have a definite probability $p(s',r'\,|\,s,a)$
which is the probability that, if the agent performs the action $a$
in the state $s$, the environment returns reward $r'$ and state
$s'$. (These $p$ functions define the MDP. Please understand
that the agent does not know $p$ while the environment does. The
agent can figure $p$ out for itself by running through the environment
many times.) 
\end{enumerate}
If we think of states, actions, and rewards all as correlated
random variables\footnote{We partially adopt the convention of using capital letters to refer
to random variables while using lowercase letters to refer to specific
values taken by those random variables.} $S_{0},A_{0},R_{1},S_{1},\ldots$, then we understand that 
\[
p(s',r'\,|\,s,a)=\text{Prob}\left(S_{t}=s',R_{t}=r'\,\big|\,S_{t-1}=s,A_{t-1}=a\right)
\]
where $t$ is some integer. In general this probability should depend
on $t$, we assume that it is actually $t$-independent which allows
$p$ to be defined. This is the Markov property.

What should an agent aspire to achieve in an MDP? An unsatisfactory approach
would be to always try to take whichever action maximizes the reward
from that action. This is a mistake because the agent may perform actions
leading to immediate rewards at the cost of poor performance in the
long term. A better
approach would be for the agent to try and select the action $a_{t}$
that gives the highest possible value of the total future reward $\sum_{t'=t}^{\infty}r_{t+1}$.
The sum of future rewards may not converge\footnote{This isn't really a problem if the environment is episodic.
This means that the state eventually ends up in a terminal state
where the process ends (like winning or losing a game). In this case
there is no convergence problem, but it remains useful, even
in these episodic cases, to make use of a discount rate.}  but that can be resolved by fixing a real number $\gamma\in(0,1)$ and defining
the return at time $t$ as 
\[
r_{t+1}+\gamma r_{t+2}+\gamma^{2}r_{t+3}+\ldots
\]
which is guaranteed to converge because $K$ is a finite set of real
numbers. The number $\gamma$ is called a discount rate. If
$\gamma$ is close to 1, then the return captures long-term success.
If $\gamma$ is close to zero, then it represents short-term success.
While setting $\gamma$ close to 1 is the ideal choice, it is often
easier to work with somewhat smaller $\gamma$.

The way that the agent influences the expectation value of the return
is by adjusting its policy. A policy is a rule that assigns
to a given state $s\in\Sigma$ a probability distribution function
over possible actions. This policy is internal to the agent and has
nothing to do with the environment. The assigned probability of selecting
action $a\in B_{s}$ given that the state is $s$ is denoted by $\pi(a\,|\,s)$.
By default, an agent selects its actions by sampling from the policy,
but during RL an agent modifies its policy to try to increase the
expectation value of the return.

\end{appendix}

\end{document}